\documentstyle[prl,twocolumn,aps]{revtex}
\input{epsf}
\begin{document}
\draft
\twocolumn[\hsize\textwidth\columnwidth\hsize\csname @twocolumnfalse\endcsname
\title{Disorder and superconductivity : a new phase of bi-particle 
localized states}

\author{Bhargavi Srinivasan and Dima L. Shepelyansky}

\address {Laboratoire de Physique Quantique, UMR 5626 du CNRS, 
Universit\'e Paul Sabatier, F-31062 Toulouse Cedex 4, France}

\date{\today}

\maketitle

\begin{abstract}
We study the two-dimensional, disordered, attractive Hubbard
model by the projector quantum Monte Carlo method and 
Bogoliubov - de Gennes mean-field theory. 
Our results for the ground state show the appearance of a new phase
with charge localization in the metallic regime
of the non-interacting model.
Contrary to the common lore, we demonstrate that mean-field theory 
fails to predict this phase and is unable to describe the 
correct physical picture in this regime. 
\end{abstract}
\vskip1pc]

\narrowtext

The transition from  superconductor to  insulator (SIT) in the presence
of disorder continues to actively attract the interest of experimentalists
and theoreticians, alike.  Experimentally, this transition has been
observed in thin-films of various materials by several 
different groups, reviewed for example in \cite{goldman}.
Theoretically, attempts were made to model the original fermionic problem 
by an effective system of interacting bosons, investigated in depth both
analytically and numerically \cite{fisher,erik}. While this
approach yields interesting results, it is important to study the
fermionic models, which directly correspond to the experimental
situation \cite{larkin}. Recently, fermion models have been studied in the
context of the SIT  by quantum Monte 
Carlo (QMC) methods \cite{scalettar} which reproduce the physics accurately.
This method yielded the transition from superconducting to
insulating behavior as a function of disorder strength. However,
several unresolved questions pertaining to
the physical  origin of this transition 
remain open. Further, a more microscopic, 
qualitative understanding of the effects of disorder and 
the properties of the localized phase
would be desirable.

For weak disorder strengths, Anderson's theorem \cite{anderson}
guarantees that the superconducting phase persists despite
disorder and the superconducting gap and other thermodynamic
properties remain unchanged. The case of stronger disorder and
non-uniform order parameters can be treated within the Bogoliubov-
de Gennes approach \cite{degennes}. Such studies have been 
carried out by several groups \cite{franz,trivedi} and the 
study for s-wave superconductors shows the persistence of a
spectral gap with relatively strong disorder \cite{trivedi}. 
In the spirit of Anderson's theorem, it is expected that
the superconducting phase penetrates the localized non-interacting
phase until the disorder strength is sufficiently large to
overcome the superconducting gap. In this framework,
the boundary between the
two phases can be estimated from the relation $\Delta \sim \Delta_1
\sim (\nu_F l^d)^{-1}$, where  $\Delta$ is the BCS superconducting gap,
$\Delta_1$ is the level spacing inside a grain of one particle  localization
length $l$ in the localized non-interacting phase and $\nu_F$ is the
density of states at the Fermi energy.
However, it is not clear whether the mean-field approximation (MFA)
remains valid in the presence of medium to strong disorder
and its validity is worth testing, by comparisons with
QMC calculations. Indeed, an indication of the failure of
the MFA comes from recent studies of the Cooper problem for
two interacting particles above a frozen Fermi sea, in the
three dimensional (3D) Anderson model (incorporating disorder) \cite{lages}.
This study 
showed that the attractive Hubbard interaction creates localized
pairs in the non-interacting metallic phase, in which the
one-particle eigenstates are delocalized. This is in contradiction
to the predictions of the MFA according to which 
Cooper pairs are delocalized in this regime.
This disagreement is expected to be even stronger in
lower dimensions (namely, the 2D case). Further, the Cooper 
approximation cannot capture all the many-body effects and
this renders interesting the study of the full, interacting, disordered
fermion model with finite density of particles.

Contrary to the expectations arising from Anderson's theorem,
we find that the attractive interaction that leads to superconductivity
gives rise to {\it a phase of bi-particle localized states (BLS) in the
metallic regime of the non-interacting disordered model}.
This confirms the indications obtained from the generalized Cooper 
problem \cite{lages} 
discussed above. Thus, our results, obtained by exact treatment 
of all many-body
quantum effects in a realistic model at finite particle density,
demonstrate convincingly the existence of the BLS phase.
Furthermore, we show that the Bogoliubov - de Gennes MFA
results are qualitatively incorrect in this phase.
The BLS phase is observed at medium disorder strength, while 
in the limit of weak disorder, the system is superconducting, as 
expected.

To investigate the interplay of interactions and disorder  
in the SIT regime, we have studied the attractive Hubbard
model with disorder, on a square lattice. The model Hamiltonian reads,
\begin{equation}
\label{hamil}
\begin{array}{l}
H = H_A + H_I  \\=
\Bigl(
 -t\sum\limits_{\langle ij \rangle, \sigma}
  \hat{a}^{\dagger}_{i,\sigma} \hat{a}_{j,\sigma}
  + \sum\limits_{i,\sigma}
  \epsilon_i \hat{a}^{\dagger}_{i,\sigma} \hat{a}_{i,\sigma}
  \Bigr)
   +  U \sum\limits_{i} \hat{n}_{i\uparrow}\hat{n}_{i\downarrow}
\end{array}
\end{equation}
   where the $\hat{a}^{\dagger}_{i,\sigma}$ ($\hat{a}_{i,\sigma}$) are the
   creation (annihilation)
   operators for a fermion of spin $\sigma$ at site $i$ with periodic
   boundary conditions, $\hat{n}_{i\sigma}$
   is the number operator for spin $\sigma$ at site $i$, $t$ is
   the hopping parameter, the Hubbard parameter, $U<0$, gives
   the strength of the screened attractive interaction and $\epsilon_i$, the
   energy of site $i$ is a random number drawn from a uniform distribution
   $[-W/2, W/2]$, which parameterizes the disorder. The first two terms
   represent the Anderson Hamiltonian and the last term represents
   the interaction $H_I$.
   The filling factor $\nu = N_p/(2 \times L^2)$, where
   $N_p$ is the number of fermions (particles) and $L$ the linear dimension
   of the system and the total number of sites is $L^2$.
It is known that the attractive Hubbard interaction induces
superconductivity; thus, we have  a useful model without 
inquiring into the physical origin of the pairing interaction.

\begin{figure}
\epsfxsize=3.4in
\epsfysize=2.6in
\epsffile{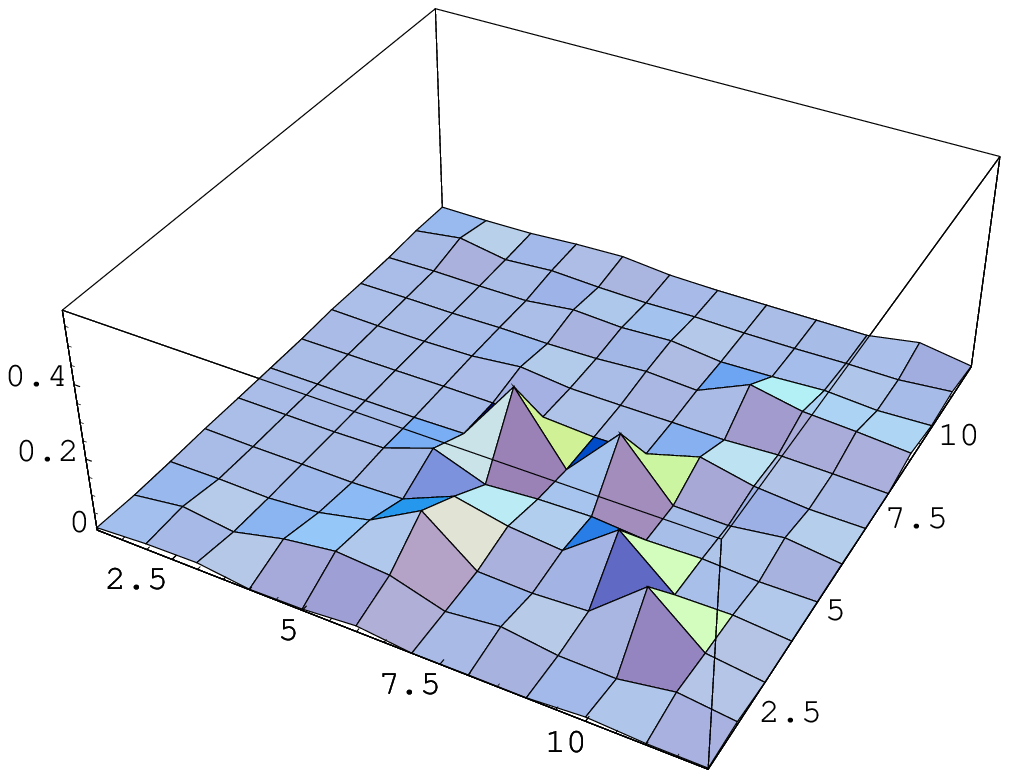}
\vglue 0.2cm
\epsfxsize=3.4in
\epsfysize=2.6in
\epsffile{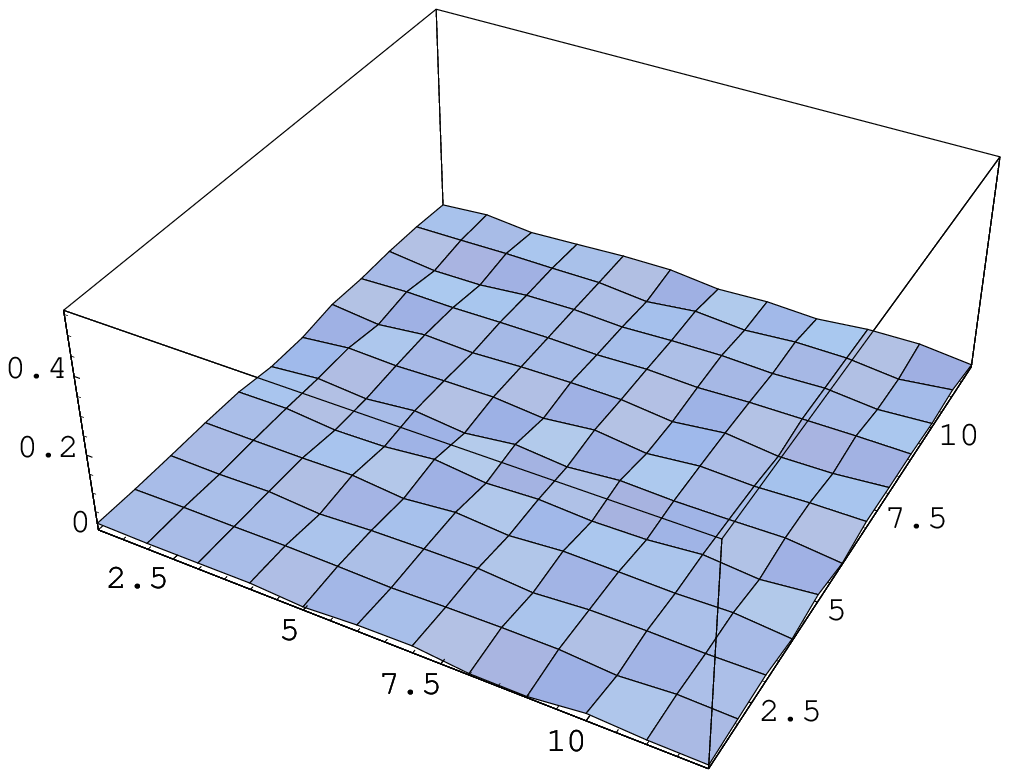}
\vglue 0.2cm
\caption{ Charge density difference, 
$\delta\rho_i$ (see text for definition),  at sites $i$ of Hamiltonian (1), 
for a $12 \times 12$ lattice with 
$W/t=5$ and $N_p = 72$ fermions for a single disorder realization. The 
upper part (a) is with $U/t= 0$ and the lower part (b)
is the result from a Bogoliubov-de Gennes calculation
with $U/t = -6.0$.
  }
\label{fig1}
\end{figure}

We have studied this model by the projector quantum Monte Carlo 
(PQMC) method, which has no fermion sign problem for $U<0$.
As is well known \cite{pqmc}, this method treats the many-body
problem exactly up to statistical errors and gives direct access
to the ground state properties of the system. The systematic error
arising from the discrete symmetric Trotter decomposition, 
of step $\Delta \tau$, is of order $(\Delta \tau)^3$.
The simulations were carried out with 
$\Delta \tau = 0.1$, with up to 60 time steps, in the $S_z = 0$ sector.
The filling factor $\nu$ was tuned around quarter filling, for different 
disorder and interaction strengths, varying the system size
$L$. The largest system studied was 144 sites with 74 particles.
We note that this model has been studied in the grand canonical 
ensemble by a finite temperature QMC method on an $8 \times 8$ lattice
with $\nu = 0.4375$ \cite{scalettar}.
The SIT was estimated in this work to occur at a 
critical value of disorder strength 
$W_c \approx 3.5t$. However, the system size dependence
of the observables was not investigated in their work.

\begin{figure}
\epsfxsize=3.4in
\epsfysize=2.6in
\epsffile{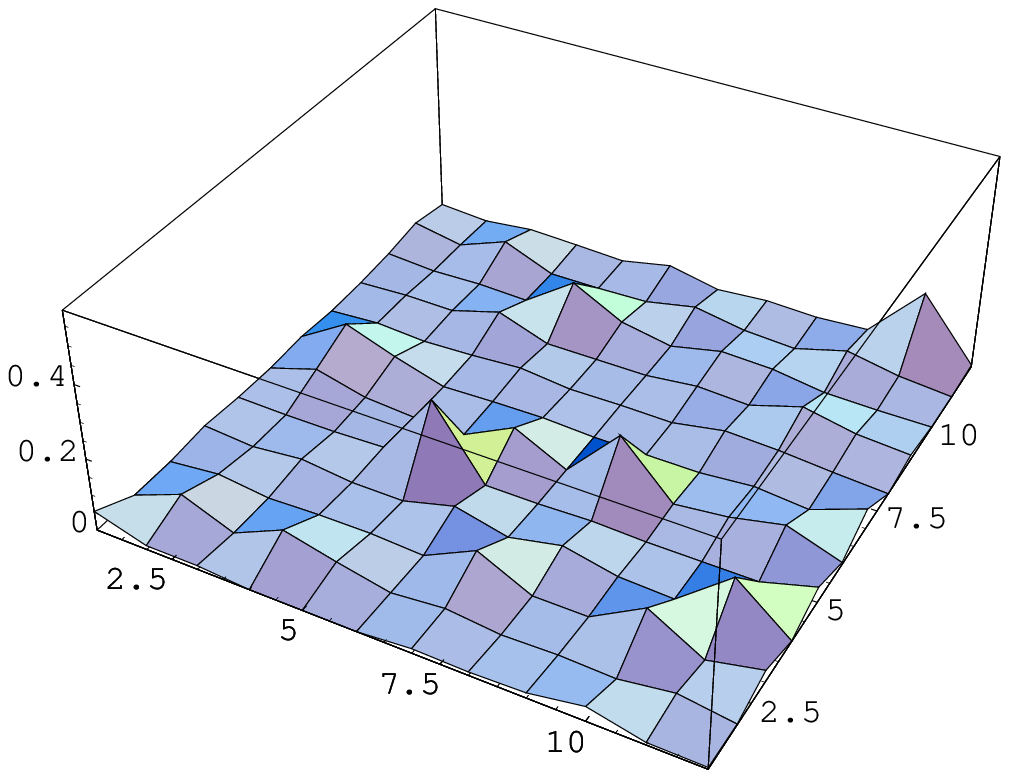}
\vglue 0.2cm
\epsfxsize=3.4in
\epsfysize=2.6in
\epsffile{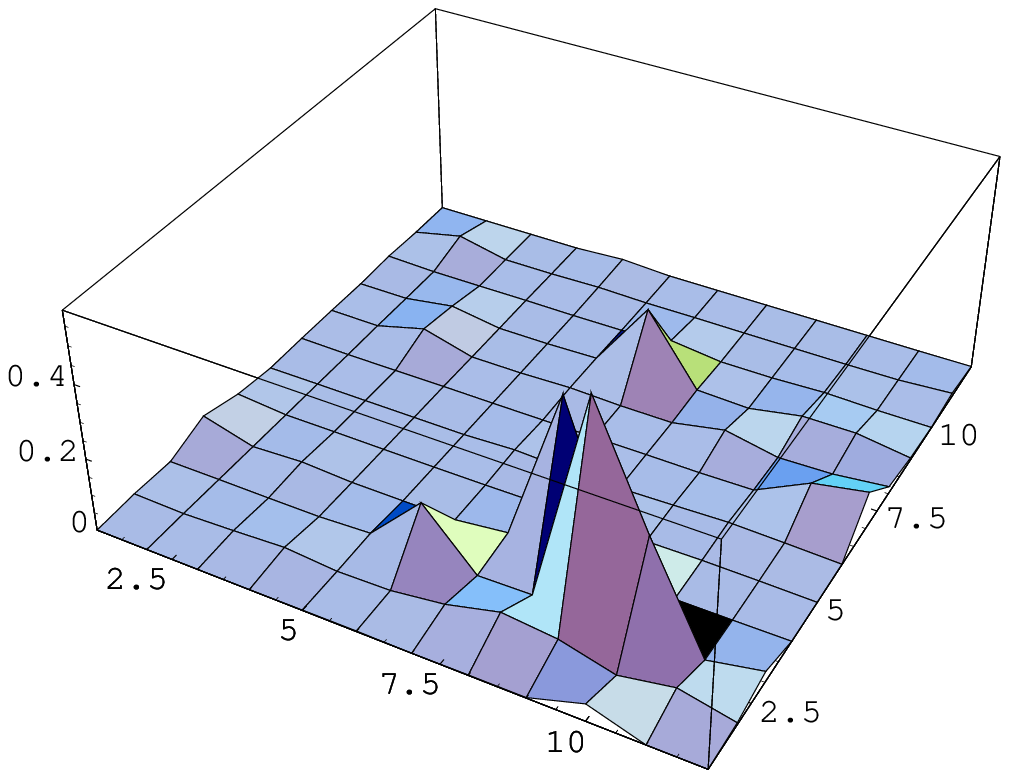}
\vglue 0.2cm
\caption{Charge density difference,
$\delta\rho_i$ at sites $i$ of Hamiltonian (1), obtained from
QMC calculations, shown here for a lattice of
$12 \times 12$ sites, with $N_p = 72$ fermions,
for the same  disorder realization as in Fig. 1. The 
upper part (a) is with $W/t=2.0$, $U/t= -6.0$ and the lower part (b)
is with $W/t=5.0$ and $U/t = -6.0$.
}
\label{fig2}
\end{figure}

To understand the physics of the model, we consider the difference
in charge density in the ground state,
$\delta \rho_i = \rho_i(N_p+2)-\rho_i(N_p)$, where
$\rho_i$ is the charge density at site $i$ and $N_p = L^2/2$ corresponds
to quarter filling ($\sum_i \delta \rho_i = 2$).
For $U = 0$, the charge density difference 
 $\delta \rho_i$ is given by the one-particle probability
 $f_i = |\psi_i|^2$ of the eigenstate at the Fermi level
 ($f_i = \delta \rho_i/2$). Due to this relation, the analysis of
 this characteristic, even in the presence of interactions, allows us to
 determine whether the charge is localized or delocalized at the Fermi level.

An example of this characteristic for a typical disorder realization
is presented in Fig. 1.  The results clearly show that the eigenstate
at the Fermi level is delocalized in the non-interacting case at the
given disorder strength $W/t = 5$. Of course, in the limit of
$L \rightarrow \infty$, eigenstates are localized for $U=0$ \cite{and79}.
Nevertheless, for the finite system sizes in Fig. 1, the localization
length is larger than the system size and the eigenstates 
correspond to a metallic regime. Certainly, it would be desirable
to study this problem in the 3D case, where the Anderson transition
clearly separates the non-interacting delocalized and localized phases.
However, it is presently too expensive, numerically,
to study comparable system sizes in 3D.
In spite of this, our data show that important physical
information can be obtained from the 2D case.

The results obtained from the Bogoliubov-de Gennes MFA for the
same disorder realization, with interaction strength $U/t = -6$
are shown in Fig. 1b. These calculations were carried out as 
described in \cite{franz,trivedi}. 
The eigenvectors and the quasiparticle excitation energies
were obtained self-consistently, with good
convergence. In addition, we  obtained the BCS value
for the order-parameter $\Delta \approx 1.36t$ in the limit of
$W/t = 0$, at $L =12$ and $U/t = -4$, as in \cite{trivedi}.
In the limit $U = 0$,  this method
reproduces $\delta \rho_i$ for finite disorder strengths $W/t$.
A comparison of Figs. 1a and 1b indicates strongly that
within the MFA, the charge density at the Fermi level is highly 
delocalized by the introduction of interactions. This
result indicates that interactions smoothen out charge fluctuations
within the MFA. Thus, at $W/t = 5$ and $U/t = -6$, the MFA gives
completely delocalized, metallic behaviour.

The PQMC results for $\delta \rho_i$ for the same disorder
realization as in Fig. 1 are displayed in Figs. 2a and 2b, 
corresponding to two
values of the disorder strength ($W/t = 2$ and $5$). 
The model parameters used are identical for Figs. 1b and 2b.
It is apparent from Fig. 2b that {\it the charge at the
Fermi energy is localized}.
Thus, the PQMC result, with a proper treatment of interactions
and disorder, differs {\it qualitatively} from the 
Bogoliubov - de Gennes mean-field calculation, which gives
the wrong physical picture. We also emphasise that for $U/t = 0$,
the charge variation $\delta \rho_i$ is delocalized (Fig. 1a)
and the pronounced, localized peak of charge in Fig. 2b is
not correlated to the charge distribution of Fig. 1a.
With decrease of disorder strength $W$, this localized charge peak
seen in Fig. 2b disappears as shown in Fig. 2a. In fact, the
charge distribution in Fig. 2a becomes closer to that of the
corresponding delocalized, non-interacting case (not shown).
These data show that the SIT takes place in the metallic non-interacting
phase.

For a more quantitative description of this unusual SIT, we 
have studied an effective inverse participation ratio (IPR), $\xi$
for an added pair of particles. The IPR is defined as
$\xi = \langle \sum_i (\delta \rho_i /2)^2 \rangle^{-1}$, where 
the brackets denote the average over 16 disorder realizations. 
For $U = 0$, this definition exactly reproduces the one-particle
IPR at the Fermi level. In the presence of interactions, this
quantity allows us to study the charge distribution of added 
pairs at the Fermi level, beyond the Cooper approximation
used in \cite{lages}. 
Physically, the IPR gives the number
of sites visited by a pair. 
Thus, the evolution of the IPR with system size
$L$, can be used to determine the transition from delocalized
to localized behavior. 

\begin{figure}
\epsfxsize=3.1in
\epsfysize=2.6in
\epsffile{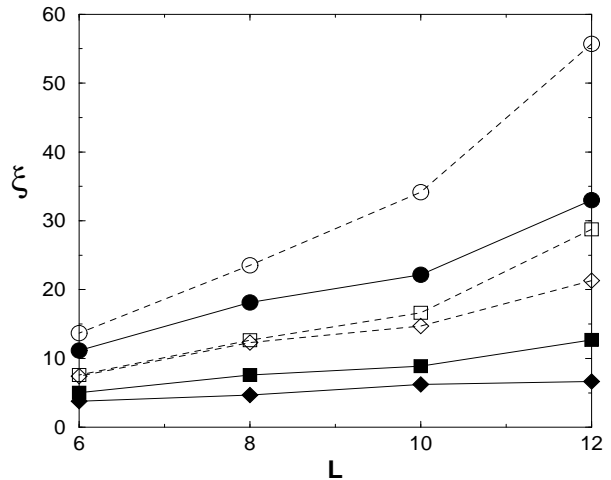}
\caption{
PQMC result for the  IPR $\xi$ obtained from the charge density distribution
 of the added pair vs. linear dimension of system $L$. Dashed lines
 are for $W/t = 2$ and full lines for $W/t = 5$, with
 $U/t$ = 0 (circles), $-4$ (squares) and $-6$ (diamonds). The
 average is carried out over 16 disorder realizations.
  }
\label{fig3}
\end{figure}

The PQMC results for the IPR as a function of system size $L$, 
disorder and interaction strengths are shown in Fig. 3.
The data clearly show two main phenomena. Firstly, 
it is clear that  interactions always
diminish the IPR, compared to the non-interacting case. 
Secondly, for moderate  disorder strengths
the pair becomes localized. Indeed, for $W/t = 5$, $\xi$
continues to grow with $L$ for $U/t = 0$ (metallic non-interacting
regime), while it remains
constant for $U/t = -6$.  For example, in the latter case,
$\xi = 6.7 \pm 0.8$ is much smaller than the total number
of sites ($L^2 = 144$), clearly indicating pair localization
induced by the attractive interaction. This is in great contrast
to the MFA result obtained as described for Fig. 1b, 
which gives complete delocalization with $\xi \approx 124$.
In fact, the $\xi$ from the MFA is significantly larger than
the non-interacting value of $\xi = 33.0$, that once more
shows that the MFA gives a physically wrong answer.
With the finite system sizes of our simulations, it is
difficult to determine the position of the SIT precisely.
However, we note that for $W/t = 2$, the system 
remains delocalized even for $U/t = -6$. Thus, we estimate that
for $U/t \simeq (-4,-6)$, the critical disorder strength is
$ W/t \simeq 4$. This is in satisfactory agreement with
the results obtained for fixed system size $L = 8$  in 
\cite{scalettar}. We note that, clearly,  well below the SIT,
at weak disorder, the Bogoliubov - de Gennes theory
becomes valid.

In fact, the appearance of this BLS phase can be understood
from the following heuristic physical argument \cite{lages}. 
The attractive Hubbard interaction strongly favors 
pair formation.  This effectively doubles the mass
of the charge carriers $m^* \propto 1/t$ and thus
enhances the effects of disorder induced localization,
which are proportion to $W/t \propto m^*$. We believe
this to be the correct physical reason for the appearance 
of this BLS phase.

In conclusion, our studies based on the PQMC method 
show the appearance of a localized phase of pairs 
which appears in the {\it metallic regime} of the 
non-interacting system. The comparison of these results
with Bogoliubov - de Gennes mean-field calculations 
show that the MFA does not predict the existence
of this phase. As the non-interacting states are
delocalized for the moderate disorder strengths 
corresponding to the BLS phase,  arguments  in
the spirit of Anderson's theorem should definitely drive the
system superconducting at the expense of the 
{\it non-interacting  localized} insulating phase.
On the contrary, our results provide evidence for the 
appearance of a new phase which is unexpected
from the accepted viewpoint. We expect that this effect
exists as well in three dimensions where it should
be much more pronounced due to the sharp Anderson
transition between the non-interacting metallic and insulating
phases.

We thank G. Benenti, G. Caldara, J. Lages and O. Sushkov
for useful discussions and remarks and IDRIS, Orsay for
access to their supercomputers.

\end{document}